\documentclass[aps,prl,reprint,showpacs,amssymb,amsmath,superscriptaddress,longbibliography]{revtex4-1}

\usepackage[T1]{fontenc}
\usepackage{graphicx}
\usepackage[usenames, dvipsnames]{xcolor}
\usepackage{times}
\usepackage{ulem}
\usepackage{verbatim}
\usepackage[separate-uncertainty=true]{siunitx}
\DeclareSIUnit{\gauss}{G}
\DeclareSIUnit{\bohrRadius}{a_0}
\usepackage{bm}

\usepackage{hyperref}
\hypersetup{%
	colorlinks,
	citecolor=blue,
	linkcolor=blue,
	urlcolor=blue
}

\newcommand{\comm}[1]{}
\newcommand{\tof}{TOF}
\newcommand{\rf}{rf}

\def\blue{\color{blue}}

\makeatletter
\def\@bibdataout@aps{%
	\immediate\write\@bibdataout{%
		@CONTROL{%
			apsrev41Control%
			\longbibliography@sw{%
				,author="08",editor="1",pages="1",title="0",year="1"%
			}{%
				,author="08",editor="1",pages="1",title="",year="1"%
			}%
		}%
	}%
	\if@filesw \immediate \write \@auxout {\string \citation {apsrev41Control}}\fi 
}
\makeatother
\begin{document}
	\title{Observation of Microcanonical Atom Number Fluctuations in a Bose-Einstein Condensate}
	
	\author{M. B. Christensen}
	\author{T.\,Vibel}
	\author{A.\,J.\,Hilliard}
	\affiliation{Center for Complex Quantum Systems, Department of Physics and Astronomy, Aarhus University, Ny Munkegade 120, DK-8000 Aarhus C, Denmark.}
	\author{M. B. Kruk}
	\author{K. Paw\l{}owski}
	\affiliation{Center for Theoretical Physics, Polish Academy of Sciences, Al. Lotnik\'{o}w 32/46, 02-668 Warsaw, Poland.}
	\author{D. Hryniuk}
	\affiliation{Center for Theoretical Physics, Polish Academy of Sciences, Al. Lotnik\'{o}w 32/46, 02-668 Warsaw, Poland.}
	\affiliation{Institute of Physics, Polish Academy of Sciences, Al. Lotnik\'{o}w 32/46, 02-668 Warsaw, Poland}
	\author{K. Rz\k{a}\.zewski}	
	\affiliation{Center for Theoretical Physics, Polish Academy of Sciences, Al. Lotnik\'{o}w 32/46, 02-668 Warsaw, Poland.}
	\author{M.\,A.\,Kristensen}
	\author{J.\,J.\,Arlt}
	\affiliation{Center for Complex Quantum Systems, Department of Physics and Astronomy, Aarhus University, Ny Munkegade 120, DK-8000 Aarhus C, Denmark.}	
	\begin{abstract}
		Quantum systems are typically characterized by the inherent fluctuation of their physical observables. Despite this fundamental importance, the investigation of the fluctuations in interacting quantum systems at finite temperature continues to pose considerable theoretical and experimental challenges. Here we report the characterization of atom number fluctuations in weakly interacting Bose-Einstein condensates. Technical fluctuations are mitigated through a combination of nondestructive detection and active stabilization of the cooling sequence. We observe fluctuations reduced by \SI{27}{\percent} below the canonical expectation for a noninteracting gas, revealing the microcanonical nature of our system. The peak fluctuations have near linear scaling with atom number $\Delta N_{0,\mathrm{p}}^2 \propto N^{1.134}$ in an experimentally accessible transition region outside the thermodynamic limit. Our experimental results thus set a benchmark for theoretical calculations under typical experimental conditions.
	\end{abstract}
	
	\maketitle
	
	% Introduction --------------------------------------------------------------------------------
	
	Bose-Einstein condensates (BECs) have become a cornerstone of current developments in quantum simulation \cite{Bloch2012}, and one would expect a complete description of their properties to be available. However, in spite of considerable theoretical effort, the atom number fluctuations between the thermal and condensed component of an interacting Bose gas at relevant atom densities are still not fully understood. 	
	In principle, such a quantum system can be described through all moments of its probability distribution, which have indeed not been obtained for large interacting BECs~\cite{Politzer1996,Navez1997,Giorgini1998,Meier1999,Kocharovsky2006}.
	To date, experiments have been limited to the first moment corresponding to the condensate fraction~\cite{Gerbier2004,Meppelink2010,Tammuz2011} as well as the critical temperature indicating the onset of condensation~\cite{Gerbier2004a,Smith2011}. Only recently BEC atom number fluctuations corresponding to the second moment have become experimentally accessible \cite{Kristensen2019}.
	
	These fluctuations pose a subtle problem with surprising challenges and rich physics \cite{Kocharovsky2006}. Historically, a description of the noninteracting Bose gas was developed within the grand canonical ensemble, which shows unphysically large fluctuations below the critical temperature \cite{Ziff1977}. This is referred to as the grand canonical catastrophe \cite{Grossmann1996,Holthaus1998}, and is one of the few examples where the predictions of different statistical ensembles differ dramatically. 
	Thus, any description of the fluctuations must be based on canonical or microcanonical approaches. 
	
	\begin{figure}[tb!]
		\centering
		\includegraphics[width=1\linewidth]{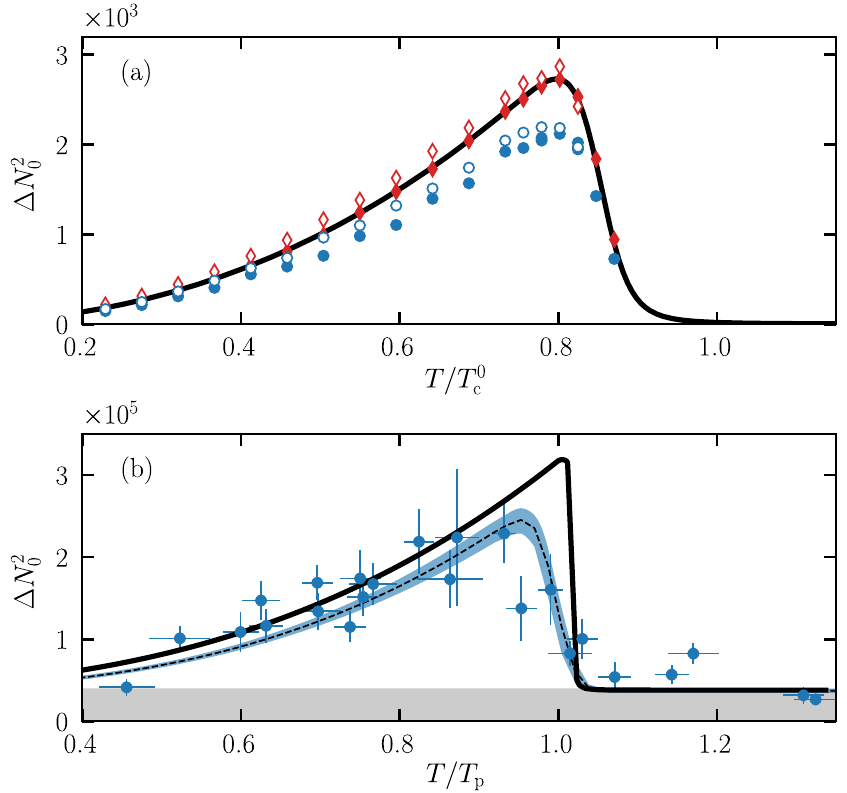}
		\caption{Atom number fluctuations of a BEC. 
			(a) Calculated variance of BEC atom number as a function of relative temperature in the canonical (red diamonds) and microcanonical (blue circles) ensembles,  with (empty symbols) and without (filled symbols) interactions at a scattering length $a = \SI{100}{\bohrRadius}$ for $N=1000$ $^{87}$Rb atoms in a trap with $\omega_a = 2\pi\times\SI{20.0}{\hertz}$ and $\lambda=10$.
			(b) Experimental variance of BEC atom number as a function of temperature (blue points). The error bars represent statistical uncertainties of the variance (vertical) and  1-$\sigma$ spread in temperature (horizontal). The dashed line  is a fit (see text), where the blue band represents the uncertainty of the fit. The gray area indicates the offset due to technical fluctuations. The aspect ratio was $\lambda = \num{7.28(2)}$ and the atom number at peak fluctuations was $N_\mathrm{p} = \num{1.49(6)e5}$. The results are plotted as a function of the temperature rescaled with the temperature at peak fluctuations. The solid black lines in both panels are exact canonical calculations for a noninteracting gas~\protect\cite{Weiss1997}.
		}
		\label{fig:dataSuppression3D}
	\end{figure}
	
	In the canonical ensemble, the asymptotic behavior of the condensate atom number variance in the noninteracting Bose gas was first obtained by Politzer~\cite{Politzer1996}. Later, exact numerical recursion relations for this noninteracting case were found~\cite{Weiss1997}, which recently allowed for an initial comparison with experimental results \cite{Kristensen2019}.
	However, the canonical ensemble assumes the existence of an external reservoir exchanging energy with the system under investigation. 
	This assumption is not met in the case of an ultracold gas isolated from the environment. Therefore, the microcanonical ensemble needs to be invoked, which is numerically demanding and currently does not allow for a full comparison at typical experimental parameters.
	
	In the microcanonical ensemble the key quantity is the partition function $\Gamma (E,\, N)$, i.e. the number of ways to distribute the energy $E$ between $N$ atoms. Calculation of $\Gamma (E,\, N)$ for a noninteracting gas in a harmonic trap is closely related to the classic mathematical problem of partitions. Discussions between Leibniz and Bernoulli inspired intensive work on this topic~\cite{Dickson1971}, with breakthroughs achieved a century ago by Ramanujan and Hardy \cite{Hardy1918}.
	In particular it was shown that the canonical and microcanonical fluctuations agree for large atom number in the noninteracting 1D gas \cite{Grossmann1997}.
	
	The results for the 3D case in the microcanonical ensemble were not known until 1997 \cite{Navez1997}, when a fourth statistical ensemble, called the Maxwell's demon ensemble, was proposed. In this ensemble the system exchanges particles with the reservoir, without changing its energy. It is applied to the thermal cloud at sufficiently low temperatures, where the BEC comprises the reservoir. 
	The ensemble yields an asymptotic expression for the microcanonical fluctuations in the noninteracting gas \footnote{The Maxwell's demon ensemble yields the variance of the number of atoms in the thermal cloud. Due to conservation of the total number of atoms, this also exactly corresponds to the variance of the number of atoms in the condensate.}
	\begin{equation}\label{eq:MicroFluctuations}
		\Delta N_0^2 = \left(\frac{\zeta(2)}{\zeta(3)} - \frac{3\zeta(3)}{4\zeta(4)}\right)N\left(\frac{T}{T_c^0}\right)^3.
	\end{equation}
	Here $T_c^0$ is the critical temperature of a noninteracting gas in a harmonic trap and $\zeta(x)$ is the Riemann zeta function. 
	In particular, the second term corresponds to the reduction in the number fluctuations in the microcanonical ensemble with respect to the canonical result given by the first term. Thus, the ratio between the canonical and microcanonical fluctuations in the 3D isotropic case is 0.39, corresponding to a reduction by \SI{61}{\percent}.

	Importantly, interactions are expected to modify the fluctuations of the condensate atom number $\Delta N_0^2$. 
	In the homogeneous case, initial results showed that interactions suppress the fluctuations by a factor of two at very low temperatures due to a strong pair correlation of the atoms reducing the number of degrees of freedom~\cite{Giorgini1998, Meier1999, Zwerger2004}.
	Moreover, interactions have been included using numerous methods such as the aforementioned Maxwell's demon ensemble~\cite{Idziaszek1999}, number-conserving quasiparticle methods~\cite{Kocharovsky2000a,Kocharovsky2000}, master-equation and hybrid quasiparticle approaches~\cite{Svidzinsky2006, Svidzinsky2010}, and a correlated many-body approach \cite{Bhattacharyya2016}. 
	However, only some of these methods are based on a microcanonical ensemble and none of them allow for the analysis of harmonically trapped interacting BECs at typical experimental atom numbers. Moreover, most approaches are only valid at low temperature and do not capture the peak fluctuations near $T_\mathrm{c}^0$.
	
	Additionally, the fluctuations are expected to scale \textit{anomalously} with atom number: $\Delta N_0^2 \propto N^{1+\gamma}$, with $ \gamma \neq 0 $~\cite{Giorgini1998,Meier1999}. This is in stark contrast to most classical systems where the central limit theorem ensures normal scaling of the fluctuations $\Delta N^2\propto N$ due to the existence of a finite microscopic coherence length scale~\cite{Zwerger2004,Yukalov2005a}. 
	The characterization of BEC atom number fluctuations is thus an important outstanding challenge that has been hindered by technical fluctuations of BEC experiments until recently. This technical limitation was mitigated in our recent experiments, enabling the current experimental characterization of BEC fluctuations.
	
	In this Letter, we characterize the atom number fluctuations in a weakly interacting BEC as a function of atom number and trapping geometry. To avoid technical fluctuations, BECs with a well-controlled atom number and temperature are prepared through a combination of nondestructive detection and active stabilization of the cooling sequence.
	We quantitatively analyze the fluctuations in two ways.
	The fluctuations are compared to an exact canonical calculation for the noninteracting Bose gas, which shows fluctuations reduced by \SI{27}{\percent} with respect to the canonical expectation, indicating the microcanonical nature of the system.
	Second, the fluctuations are fitted with a phenomenological model showing anomalous atom number scaling with an exponent $\gamma = \num{0.134(5)}$, which we compare to the expectation in the thermodynamic limit. 
	
	\begin{figure}[b]
		\centering
		\includegraphics[width=1\linewidth]{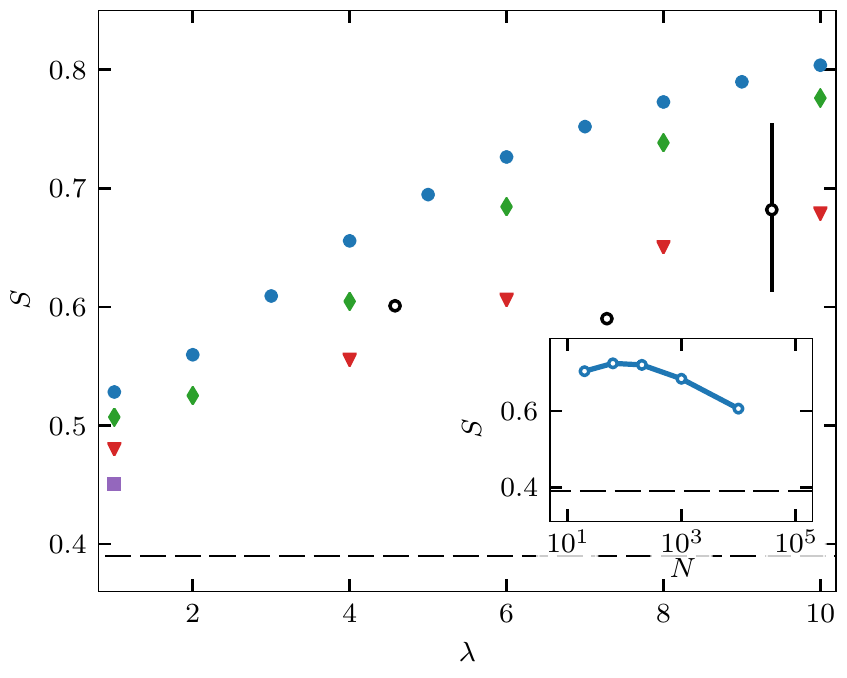}
		\caption{Ratio, $S$, between the calculated peak variance of the noninteracting BEC in the microcanonical and canonical ensembles as a function of the trap  aspect ratio $\lambda$ for $N=10^2$ (blue points), $N=10^3$ (green diamonds),  $N=10^4$ (red triangles) and $N=10^5$ (purple square) atoms. The black dashed line corresponds to the asymptotic value estimated within the Maxwell's demon ensemble (see Eq.\eqref{eq:MicroFluctuations}). The three experimental points with the lowest atom number $N_\mathrm{p} = \num{8.8e4} - \num{10.4e4}$ are shown for comparison (open points). The typical error bar shown indicates the \SI{68.2}{\percent} confidence interval on the experimental value. (Inset) $S$ as a function of the number of atoms for fixed aspect ratio $\lambda=6$.}
		\label{fig:ARandNtheory}
	\end{figure}
	
	In a first step, we theoretically estimate the importance of using an interacting microcanonical ensemble with respect to the more readily available canonical ensemble calculations. This is achieved for ensembles of up to $10^5$ atoms using a new computational method.
	In brief, we apply the usual Metropolis algorithm but perform a random walk in the space of Fock states, accounting for Bose enhancement with perturbatively included interactions. The random walk results in the set of visited Fock states obeying canonical ensemble statistics. By post-selecting samples according to their energy, we obtain representative ``microstates'' of the microcanonical ensemble~\cite{SupplementaryMaterial}. Figure~\ref{fig:dataSuppression3D}(a) show the results for $N=1000$ atoms for trap aspect ratio $\lambda = 10$. 
	Importantly, the microcanonical result lies clearly below the canonical expectation as indicated by Eq.~\eqref{eq:MicroFluctuations}. In the elongated case shown here, the ratio, $S$, between the peak variance in the microcanonical and canonical ensembles is $S = 0.72$, corresponding to a reduction by \SI{28}{\percent} for this atom number. Overall, it is evident from Fig.~\ref{fig:dataSuppression3D}(a) that the effect of interactions is small and leads to a minor increase of the variance in both ensembles.
	
	Given the small effect of interactions, we calculate $S$ for a range of aspect ratios and atom numbers for the noninteracting case as shown in Fig.~\ref{fig:ARandNtheory}. For low aspect ratios $\lambda$ and large atom numbers the ratio $S$ tends towards the limiting value 0.39 given by Eq.~\eqref{eq:MicroFluctuations}. In highly elongated traps with large aspect ratios, $S$ tends towards 1, since microcanonical and canonical fluctuations are identical in the noninteracting 1D case.
	The dependence of $S$ on the atom number is shown in Fig.~\ref{fig:ARandNtheory} (inset) for $\lambda = 6$. This indicates that the microcanonical and canonical results scale differently with atom number. In particular we observe a roughly logarithmic dependence of $S$ on $N$ for atom numbers $10^2 < N < 10^4$.
	These results show that a clear reduction of the fluctuations below the canonical result is expected under realistic conditions and motivates the following experimental investigation.
	
	% Experimental method --------------------------------------------------------------------------------
	The experimental apparatus used to produce BECs was previously described in detail in Ref.~\cite{Kristensen2017}. Briefly, around $10^9$ $^{\text{87}}$Rb~atoms are first captured and cooled in a magneto-optical trap. The cloud is then optically pumped to the $\left|F=2, m_F =2\right\rangle $ state. Then it is transported into a cigar-shaped quadrupole-Ioffe-configuration magnetic trap with axial trapping frequency $\omega_a$ and radial frequency $\omega_\rho$~\cite{SupplementaryMaterial}, where it is cooled by radio-frequency (\rf{}) forced evaporation.
	
	The sequence outlined above is typically subject to technical fluctuations that prevent the observation of atom number fluctuations. To overcome this challenge, a stabilization procedure is initiated when the cloud contains $\sim 4\times 10^6$ atoms at a temperature $~\SI{14}{\micro\kelvin}$. The cloud is probed using minimally destructive Faraday imaging, and the atom number is corrected using a weak \rf{} pulse of controllable duration. The pulse removes excess atoms, and the outcome of the stabilization procedure is verified by a second Faraday measurement. This allows us to prepare a well-controlled number of atoms with a relative stability at the $10^{-4}$ level~\cite{Gajdacz2016, Kristensen2017}.
	
	For efficient cooling towards BEC, a tight magnetic trap and consequently large collision rate is desirable. In its most compressed configuration, our trap has an aspect ratio $\lambda = 17$. This can lead to significant phase fluctuations across the spatial extent of the cloud~\cite{Hellweg2003}, which evolve into density modulations during time of flight (\tof{}) and thus hinder the precise determination of atom number and temperature. 
	We therefore decompress the magnetic trap for our measurements, limiting the aspect ratio to $4.5 < \lambda < 10$ where the phase coherence length is larger than the condensate length in the long direction.
	
	In the final step of the experimental sequence, BECs are produced by forced evaporation ending at an \rf{}-frequency corresponding to the desired BEC temperature. To ensure proper thermal equilibrium, the BEC is first held at the final \rf{} frequency for {$\SI{800}{\milli\second}$\blue} and a further \SI{400}{\milli\second} without \rf{} radiation before the trap is turned off.
	
	The clouds are probed after $\SI{35}{\milli\second}$ \tof{} expansion using resonant absorption imaging on the $F=2 \rightarrow F'=3 $ cycling transition. 
	Fringes in the processed image due to vibrations of optical components are mitigated  by minimizing the time-delay between the absorption and reference image. Optical pumping is applied between these images to transfer the atoms to the transparent $F=1$ state. Thus, the atom and beam images can be taken only $\SI{340}{\micro\second}$ apart, limited only by the camera shift speed. 
	
	For each experimental configuration corresponding to a chosen aspect ratio and atom number, the fluctuations were measured as a function of the temperature $T$. 
	At every temperature (corresponding to an \rf{} end frequency) a set of 60 measurements according to the experimental sequence outline above was taken~\footnote{Measurements for which the stabilization process failed were excluded, such that each set includes at least 45 measurements.}.  
	
	\begin{figure}[b]
		\centering
		\includegraphics[width=1\linewidth]{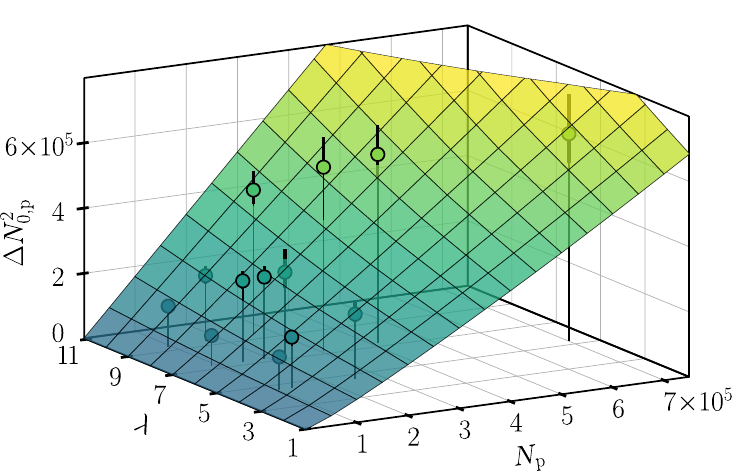}
		\caption{Peak fluctuations of a BEC. (Data points) Observed peak variance of the BEC atom number as a function of the atom number at peak variance $N_\mathrm{p}$ and trap aspect ratio $\lambda$.  The thick black bars indicate the \SI{68.2}{\percent} confidence interval. Narrow stems connect the data points to the base to guide the eye. (Surface) Theoretical expectation for a noninteraction canonical ensemble scaled according to Eq.~\protect\eqref{eq:SuppressionModel}.
		}
		\label{fig:SuppressionFactor}
	\end{figure}
		
	% Data evaluation method --------------------------------------------------------------------
	To determine the BEC atom number in each image, the wings of the cloud are fit with a Bose-enhanced thermal distribution from which the temperature is obtained. The fitted distribution is subtracted from the image, and the BEC atom number $N_0$ is obtained by integration of the remaining density.
	However, the variance cannot be determined directly from $N_0$, since small remaining drifts of the magnetic offset field lead to minor temperature variations with a median standard deviation of $\sim \SI{3}{\nano\kelvin}$. We eliminate this drift by subtracting a linear fit of the condensate number as a function of the total atom number~\cite{Kristensen2019} and determine the residuals $\eta_i$, where $i$ indicates the order in time.
	The BEC atom number variance is then given by a two-sample variance of the residuals
	\begin{equation}\label{eq:TwoSampleVariance}
		\Delta N_0^2 = \frac{1}{2}\left\langle \left(\eta_{i+1} - \eta_i\right)^2 \right\rangle.
	\end{equation}
	Thus, this two-sample variance contains the BEC fluctuations and detection noise, but excludes slow technical drifts.
	Figure~\ref{fig:dataSuppression3D}(b) shows this variance of the measurements as a function of $T$ for a given configuration of the experiment.
	
	The peak fluctuations are determined by a fit to $\Delta N_0^2$ also illustrated in Fig.~\ref{fig:dataSuppression3D}(b). The fit model is inspired by the asymptotic behavior of the fluctuations in a noninteracting gas $ \Delta N_0^2 = \zeta(2) \left(\frac{k_\mathrm{B}}{\hbar\overline{\omega}}\right)^3T^3$, where $\overline{\omega}$ is the geometric mean of the trapping frequencies~\cite{Politzer1996}.
	Moreover, the fluctuations decay in near steplike manner close to the critical temperature, which we model with a Heaviside step function $\Theta(T_p - T)$, where $T_\mathrm{p}$ is the temperature at the peak fluctuations. 
	To account for small temperature drifts the expression is furthermore convolved with a normal distribution $\mathcal{N}(T,\sigma_T)$ centered on the temperature $T$ with a standard deviation $\sigma_T$ given by the median of the measured temperature variation. Thus, we fit the data with the model
	\begin{equation}\label{eq:fitModel}
		\Delta N_0^2(T) = \left(f\ast g\right)(T) +\mathcal{O}
	\end{equation}
	where $f$ and $g$ are given by
	\begin{align}
		f(T) &= \Delta N_\mathrm{0,p}^2 \left(\frac{T}{T_\mathrm{p}}\right)^3 \Theta(T_\mathrm{p}-T),
		\\
		g(T) &= \mathcal{N}(T,\sigma_T).
	\end{align}
	The three fit parameters are the peak atom number variance $\Delta N_{\mathrm{0,p}}^2$, $T_\mathrm{p}$ and an offset $\mathcal{O}$ which accounts for experimental noise~\cite{Kristensen2019}. Since the atom number varies as a consequence of the evaporation, we determine the atom number at peak fluctuations $N_\mathrm{p}$ from a spline interpolation of the atom number as a function of the temperature for all measurements in a given configuration, evaluated at $T_\mathrm{p}$.
	Figure~\ref{fig:SuppressionFactor} shows the measured peak atom number variance for all 13 experimental configurations of aspect ratio and atom number (for a tabular presentation of the data see Supplemental Material~\cite{SupplementaryMaterial}). The three points at lowest atom number are also included in Fig.~\ref{fig:ARandNtheory}.
		
	% Main result -------------------------------------------------------------------------------
	We quantitatively evaluate the peak fluctuations using two approaches.
	First, we compare the measured fluctuations to the expected fluctuations in a canonical noninteracting gas, which can be calculated under conditions comparable to the experiment.
	Second, the fluctuations are investigated using a phenomenological model to probe their dependence on atom number and trapping geometry.
	
	\begin{figure}[b]
		\centering
		\includegraphics[width=1\linewidth]{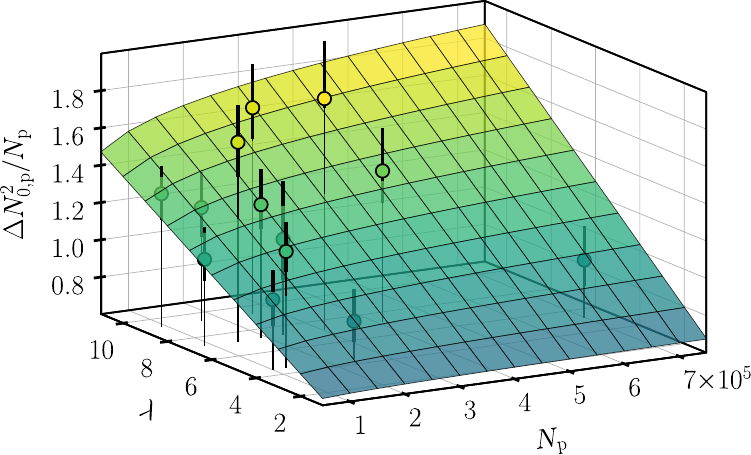}
		\caption{Scaling model for peak atom number fluctuations of a BEC. (Data points) Normalized peak variance of the BEC atom number as a function of $ N_\mathrm{p} $ and aspect ratio $ \lambda $. The thick black bars indicate the \SI{68.2}{\percent} confidence interval. Narrow stems connect the data points to the base to guide the eye. (Surface) Phenomenological model given by Eq.~\protect\eqref{eq:AnalyticFitFunction} of the fluctuations.
		}
		\label{fig:AnalyticDependencies}
	\end{figure}
	
	To compare experiment and theory the average ratio $S$ between the observed fluctuations and the exact theoretical result for the noninteracting canonical ensemble $\Delta N_{\mathrm{0,p,can}}^2$ is evaluated. We obtain $S$ from a fit of 
	\begin{equation}\label{eq:SuppressionModel}
		\Delta N_{\mathrm{0,p}}^2 = S(1 + a\,\delta \lambda + b\,\delta\!\ln(N))\Delta N_{\mathrm{0,p,can}}^2,
	\end{equation}
	inspired by the scalings with atom number and aspect ratio in Fig.~2. The coefficients $a$ and $b$ allow for a linear scaling with the trap aspect ratio and the logarithm of the atom number, where $\delta x = x - \langle x \rangle$ denotes that the mean over the set of experimental parameters has been subtracted.
	This yields an average ratio $S=\num{0.73(5)}$ corresponding to a $\SI{27}{\percent}$ reduction of the fluctuations and clearly reveals the microcanonical nature of our system with reduced fluctuations below the canonical ensemble expectation.
	The coefficients $a = \num{0.04(4)}$ and $b = \num{0.24(26)}$ represent a correction to $S$ which are barely resolved within our data range, supporting the interpretation that the microcanonical nature of the system constitutes the most important deviation from the canonical result.

	We use a complementary approach to analyze the scaling of the fluctuations with atom number and aspect ratio $\lambda$. 
	Inspired by Eq.~\eqref{eq:MicroFluctuations}, we use the phenomenological model
	\begin{equation}\label{eq:AnalyticFitFunction}
		\Delta N_{0,\mathrm{p}}^2/N = a + b\lambda N^\gamma,
	\end{equation}
	where $a$, $b$ and $\gamma$ are fit coefficients allowing for linear aspect ratio and a nonlinear number dependence. In particular, the exponent $\gamma$ quantifies the degree of anomalous atom number scaling. 
	
	We first fit the exact canonical result $ \Delta N_{\mathrm{0,p,can}}^2 $ with this model which yields $a = \num{1.327(14)}$, $b = \num{1.5(5)}$ and $\gamma = \num{-0.27(3)}$. Thus, the coefficient $ a $ is close to its limiting value $ \zeta(2)/\zeta(3) = 1.37 $ corresponding to the first term in Eq.~\ref{eq:MicroFluctuations}. More significantly, the negative sign of $ \gamma $ combined with the relatively large value of $ b $ show that even the exact theoretical result scales anomalously with $ N $. This can be attributed to the transition from a system with reduced dimensionality at low atom numbers to the limiting 3D case~\cite{Kristensen2019}. Hence, we do not expect our experiment to reflect the thermodynamic limit, where the scaling has been investigated theoretically~\cite{Giorgini1998,Idziaszek1999, Meier1999}.
	
	Figure~\ref{fig:AnalyticDependencies} shows the experimental data and the fit according to Eq.~\eqref{eq:AnalyticFitFunction} using a maximum likelihood method. Contrary to the noninteracting theoretical case, we find anomalous scaling with $\gamma = \num{0.134(5)} > 0$.
	We interpret our result as being due to the interplay between the dimensionality effects in a microcanonical system and interaction effects, which are predicted to yield $ \gamma=1/3 $ in the thermodynamic limit~\cite{Giorgini1998,Meier1999}. In particular, Fig.~\ref{fig:ARandNtheory} shows that a microcanonical system generally scales different from a canonical one in both $ \lambda $ and $ N $.
	Since $\gamma > 0$, Eq.~\eqref{eq:AnalyticFitFunction} diverges and the coefficient $a = \num{0.55(4)}$ cannot be interpreted as a limiting value. Moreover, $b = \num{0.020(12)}$ is very small due to the large value of $N^\gamma$. This result thus provides a simple analytical model as a benchmark for future theoretical investigations.

	% Conclusion --------------------------------------------------------------------------------
		In conclusion, we have measured the peak fluctuations of a weakly interacting Bose-Einstein condensate as a function of atom number and trap aspect ratio.  The significant differences between a canonical and microcanonical ensemble description of these fluctuations were investigated numerically by post-processing  ensembles generated via the quantum Metropolis algorithm. Experimentally the fluctuations were found to be reduced by \SI{27}{\percent} relative to the expectation for a canonical    gas, in qualitative agreement with our theoretical prediction. This clearly demonstrates the micro-canonical nature of the system and thus guides future theoretical work.
	In addition the peak fluctuations were found to follow a weakly anomalous scaling with the total atom number $\Delta N_{0p}^2 \propto N^{1.134}$. We interpret the scaling as an interplay between dimensional effects and the limiting case in the thermodynamic limit with $\Delta N_{0p}^2 \propto N^{4/3}$.
	
	Future experiments will aim to observe the slight increase of the fluctuations due to interactions, predicted by our theoretical calculation. To date, this effect is shrouded by measurement uncertainties and the dependence of the fluctuations on trapping geometry. Moreover, the use of larger atomic clouds will provide experimental insight in the fluctuations in the thermodynamic limit~\cite{Giorgini1998, Meier1999}. Finally, technical improvements will allow for larger data sample sizes and enable the investigation of the time dependence and higher moments of the atom number distribution~\cite{Zeiher2020}.

	% Acknowledgements --------------------------------------------------------------------------------
	\begin{acknowledgments}
		M.B.C., T.V., A.J.H., M.A.K., and J.J.A. acknowledge support by the Villum Foundation, the Carlsberg Foundation, and the Danish National Research Foundation through the Center of Excellence ``CCQ'' (Grant agreement No. DNRF156).
		K.R., M.B.K. and D.H. acknowledge support from the (Polish) National Science Center Grant No. 2018/29/B/ST2/01308.
		K.P. acknowledges support from the (Polish) National Science Center Grant No. 2019/34/E/ST2/00289.
		This research was supported in part by PLGrid Infrastructure.
		Center for Theoretical Physics of the Polish Academy of Sciences is a member of the National Laboratory of Atomic, Molecular and Optical Physics (KL FAMO). MBK acknowledges support from the (Polish) National Science Center Grant No. 2018/31/B/ST2/01871. This article has been supported by the Polish National Agency for Academic Exchange under Grant No. PPI/ PZA/2019/1/00094/U/00001.
	\end{acknowledgments}	
	\normalem % Reset emphasis style to allow line breaks in book titles

\end{document}

\title{Supplemental Material to ``Observation of Microcanonical Atom Number Fluctuations in a Bose-Einstein Condensate''}
\author{M. B. Christensen}
\author{T.\,Vibel}
\author{A.\,J.\,Hilliard}
\affiliation{Center for Complex Quantum Systems, Department of Physics and Astronomy, Aarhus University, Ny Munkegade 120, DK-8000 Aarhus C, Denmark.}
\author{M. B. Kruk}
\author{K. Paw\l{}owski}
\affiliation{Center for Theoretical Physics, Polish Academy of Sciences, Al. Lotnik\'{o}w 32/46, 02-668 Warsaw, Poland.}
\author{D. Hryniuk}
\affiliation{Center for Theoretical Physics, Polish Academy of Sciences, Al. Lotnik\'{o}w 32/46, 02-668 Warsaw, Poland.}
\affiliation{Institute of Physics, Polish Academy of Sciences, Al. Lotnik\'{o}w 32/46, 02-668 Warsaw, Poland}
\author{K. Rz\k{a}\.zewski}
\affiliation{Center for Theoretical Physics, Polish Academy of Sciences, Al. Lotnik\'{o}w 32/46, 02-668 Warsaw, Poland.}
\author{M.\,A.\,Kristensen}
\author{J.\,J.\,Arlt}
\affiliation{Center for Complex Quantum Systems, Department of Physics and Astronomy, Aarhus University, Ny Munkegade 120, DK-8000 Aarhus C, Denmark.}

\maketitle

\section{Tabulated experimental values}
Table~\ref{tbl:ExperimentalValues} provides the experimental parameters and the values obtained from fits of Eq.~(3) in the main text, ordered according to atom number at peak fluctuations $N_\mathrm{p}$. Additionally, the theoretical value for the fluctuations within the canonical ensemble is given. The data corresponds to Fig.~3 and Fig.~4 in the main text.

\begin{table}[h]
	\caption{Comparison of the measured fluctuations $\Delta N_0^2$ and the theoretical value within the canonical ensemble $\Delta N_{0,\mathrm{can}}^2$. In addition the atom number $N_\mathrm{p}$ and temperature $T_\mathrm{p}$ at the peak fluctuations as well as the axial and radial trapping frequencies $\omega_a$ and $\omega_\rho$, the trap aspect ratio $\lambda = \omega_\rho/\omega_a$ and the fitted offset $\mathcal{O}$ are given.}
	
	\label{tbl:ExperimentalValues}
	\begin{tabularx}{1\linewidth}{d{1.7} X D{x}{}{3.5} X d{2.2}  X d{3.2}  X d{1.2}  X d{1.9} X d{1.7} X d{1.2}}
		\toprule
		\mc{$N_\mathrm{p}/10^5$}\footnotemark[1] && \mc{$T_\mathrm{p}$ (\si{\nano\kelvin})}\footnotemark[1]\footnotemark[2] && \mc{$\omega_a/2\pi$ (\si{\hertz})} && \mc{$\omega_\rho/2\pi$ (\si{\hertz})} && \mc{$\lambda$} && \mc{$\mathcal{O}/10^5$ \footnotemark[1]} && \mc{$\Delta N_0^2/10^5$}\footnotemark[1] && \mc{$\Delta N_{0,\mathrm{can}}^2/10^5$} 
		\footnotetext{The quoted uncertainties are statistical bounds on the fitted values corresponding to a \SI{68.2}{\percent} confidence interval.}
		\footnotetext{The quoted uncertainties are purely statistical and additional systematic uncertainties due to finite switching time and non-ballistic expansion during the time of flight are not included~\protect\cite{Gerbier2004}. }
		\\		
		\midrule
		0.89(5) 	&& 123x^{+6}_{-3} 	&& 17.56 	&& 127.80 	&& 7.28 	&& 0.233^{+0.012}_{-0.012} 	&& 0.94^{+0.15}_{-0.09} 	&& 1.59 \\[1pt]
		0.95(19) 	&& 148x.3^{+2.9}_{-0.9} 	&& 17.68 	&& 165.84 	&& 9.38 	&& 0.347^{+0.019}_{-0.019} 	&& 1.24^{+0.13}_{-0.13} 	&& 1.82 \\[1pt]
		1.04(12) 	&& 93x.2^{+0.7}_{-0.9} 	&& 17.10 	&& 78.28 	&& 4.58 	&& 0.37^{+0.03}_{-0.03} 	&& 1.01^{+0.17}_{-0.13} 	&& 1.68 \\[1pt]
		1.28(7) 	&& 95x.2^{+5.9}_{-0.3} 	&& 17.10 	&& 78.28 	&& 4.58 	&& 0.45^{+0.03}_{-0.03} 	&& 1.56^{+0.19}_{-0.13} 	&& 2.04 \\[1pt]
		1.49(6) 	&& 148x.4^{+0.4}_{-7.4} 	&& 17.56 	&& 127.80 	&& 7.28 	&& 0.38^{+0.02}_{-0.02} 	&& 2.5^{+0.3}_{-0.3} 	&& 2.54 \\[1pt]
		1.68(2) 	&& 176x.5^{+2.4}_{-0.9} 	&& 17.68 	&& 165.84 	&& 9.38 	&& 0.52^{+0.03}_{-0.03} 	&& 2.0^{+0.3}_{-0.3} 	&& 3.10 \\[1pt]
		1.915(7) 	&& 159x.8^{+0.4}_{-0.3} 	&& 17.56 	&& 127.80 	&& 7.28 	&& 0.46^{+0.03}_{-0.03} 	&& 2.5^{+0.3}_{-0.2} 	&& 3.26 \\[1pt]
		2.32(3) 	&& 169x.0^{+2.7}_{-1.1} 	&& 17.56 	&& 127.80 	&& 7.28 	&& 0.72^{+0.03}_{-0.03} 	&& 2.6^{+0.7}_{-0.5} 	&& 3.94 \\[1pt]
		2.52(10) 	&& 123x.7^{+7.2}_{-0.5} 	&& 17.10 	&& 78.28 	&& 4.58 	&& 0.89^{+0.04}_{-0.04} 	&& 2.0^{+0.4}_{-0.3} 	&& 3.97 \\[1pt]
		2.61(7) 	&& 199x.9^{+5.7}_{-1.0} 	&& 17.68 	&& 165.84 	&& 9.38 	&& 0.77^{+0.03}_{-0.03} 	&& 4.4^{+0.6}_{-0.4} 	&& 4.57 \\[1pt]
		3.1(10) 	&& 183x.3^{+8.0}_{-1.2} 	&& 17.56 	&& 127.80 	&& 7.28 	&& 0.97^{+0.06}_{-0.06} 	&& 5.64^{+0.91}_{-0.12} 	&& 5.01 \\[1pt]
		4.1(17) 	&& 211x^{+3}_{-11} 	&& 17.56 	&& 127.80 	&& 7.28 	&& 1.52^{+0.11}_{-0.11} 	&& 5.8^{+0.9}_{-0.7} 	&& 6.73 \\[1pt]
		7.03(6) 	&& 212x.7^{+1.8}_{-1.6} 	&& 17.29 	&& 92.52 	&& 5.35 	&& 2.33^{+0.14}_{-0.14} 	&& 6.3^{+1.2}_{-0.9} 	&& 10.84 \\[1pt]
		
		\bottomrule
	\end{tabularx}
\end{table}

\section{Numerical method}
In the following we briefly present the algorithm used to compute the atom number fluctuations in a Bose-Einstein condensate presented in the main text. Calculations of the atom number fluctuations in the non-interacting canonical ensemble were performed using exact numerical recursion relations~\cite{Weiss1997}. Our algorithm for the calculation of fluctuations including interactions and within the microcanonical ensemble is outlined below.

%\vspace{3mm}
\textit{Statistical ensemble:}
We invoke the interpretation of a thermal state as an ensemble of many configurations of a given system, where a given configuration corresponds to a specific distribution of the atoms among the available single particle states.
The algorithm aims to generate $M$ representative members of a statistical ensemble in order to estimate the condensate occupation and fluctuations according to
\begin{equation}
	\langle N_0 \rangle\approx  \frac{1}{M}\sum_{i=1}^M N_0^{(i)},\quad\langle N_0^2 \rangle \approx  \frac{1}{M}\sum_{i=1}^M \left(N_0^{(i)}\right)^2 \quad\Delta N^2_0  = \langle N_0^2 \rangle-\langle N_0 \rangle^2,
	\label{eq:N0-numerically}
\end{equation}
where the index $i$ runs over the chosen configurations, and $N_0^{(i)}$ is the number of condensed atoms in the $i$-th configuration.
The representative members are collected during a random walk in the space of system configurations using the Metropolis algorithm.

%\vspace{3mm}
\textit{Single configuration and its energy:}
We represent a single configuration of the system as a Fock state 
$|n_0, \, n_1,\,\ldots n_{k_{\rm max}}\rangle_i$. Here $n_k$ denotes the occupation of the $k$-th energy level. To simulate the canonical ensemble we limit the space of Fock states to those obeying the constraint $\sum_{k=0}^{k_\mathrm{max}} n_k = N$, where $N$ is the total number of atoms. 
The energy of a single copy is given by
\begin{equation}
	E_i = \sum_{k=0}^{k_{\rm max}}\,\epsilon_k\, (n_k)_i ,
\end{equation}
where $\epsilon_k$ is the $k$-th single-particle energy, and we use index $i$ in the symbol $(n_k)_i$ to stress that $n_k$ is the occupation of the $k$-th energy level in the $i$-th configuration of the system. 
Our results do not depend on a cut-off $k_{\rm max}$ corresponding to the maximal single particle state {taken into account} in the simulation, provided that it is sufficiently high to cover all important physical phenomena.

%\vspace{3mm}
\textit{Algorithm:}
We use the Metropolis algorithm, which in essence consists of the following steps:
\begin{itemize}
	\item[1)] We initiate the first state $|\bm{n}\rangle_1=|n_0, \, n_1,\,\ldots n_{k_{\rm max}}\rangle_1$ by randomly choosing the $n_i$'s.
	The total energy of this configuration is $E_{1} = \sum_{k=0}^{k_{\rm max}}\, \epsilon_k \,(n_k)_1$.
	
	\item[2)] Next, a new candidate state is generated from $|\bm{n}\rangle_1$ by shifting a particle from a randomly chosen energy level $k_{0}$ to another randomly chosen energy level $k_{1}$. 
	The energy of the new {candidate} vector of occupations $|\bm{n}\rangle_{\rm cand}$ is
	\begin{equation}
		E_{\rm cand} = \sum_{k=0}^{k_{\rm max}} \epsilon_k\,(n_k)_1  - \epsilon_{k_0}+\epsilon_{k_1}.
	\end{equation} 
	
	\item[3)]
	If $E_{\rm cand}<E_{1}$ then $|\bm{n}\rangle_{\rm cand}$ is accepted as $|\bm{n}\rangle_2$. Otherwise, the candidate state is accepted with probability $e^{ \left(E_{\rm cand} - E_1\right)/k_B T}$, where $k_B$ is the Boltzmann constant.
	If the candidate is rejected we set $|\bm{n}\rangle_2\coloneqq |\bm{n}\rangle_1$.
	
	\item[4)] We repeat steps 2) and 3) to generate $|\bm{n}\rangle_3, |\bm{n}\rangle_4\ldots |\bm{n}\rangle_m \ldots $
\end{itemize}
During the first iterations, the sampled configurations tend towards attractors as in a thermalization process. The samples during this burn-in period are discarded, and we only record states from a random walk around equilibrium. 
The remaining  set of visited Fock states is our representation of the canonical ensemble of the system, and it is used to compute the average occupation and fluctuations of the condensate atom number according to Eq.~\eqref{eq:N0-numerically}.

%\vspace{3mm}
\textit{Microcanonical ensemble:}
The microcanonical ensemble is an ensemble of states with constant atom number $N$ and energy $E$. To mimic the microcanoncial ensemble we post process the sampled states by removing configurations with a total energy $E_i$ outside a narrow energy window around the most probable energy. When the window is narrowed, the averages $\langle N_0 \rangle$ and $\Delta^2 N_0$ stabilize. The remaining configurations are our representative members of the microcanonical ensemble.

%\vspace{3mm}
\textit{Interactions:}
To account for interactions we modify the formula for energy using perturbation theory
\begin{equation}
	E_i^{\rm int} = E_i + {}_i\langle n_0, \, n_1,\,\ldots n_{k_{\rm max}}| \hat{H}_{\rm int}  |n_0, \, n_1,\,\ldots n_{k_{\rm max}}\rangle_i,
\end{equation}
where $\hat{H}_{\rm int}$ is the part of the Hamiltonian that describes the interaction energy. The results from the main text correspond to
\begin{equation}
	\hat{H}_{\rm int} = \int d^3 r \hat{\Psi}^{\dagger} (\bm{r})\hat{\Psi}^{\dagger}(\bm{r})\hat{\Psi}(\bm{r})\hat{\Psi} (\bm{r}),
\end{equation}
where $\hat{\Psi}(\bm{r})$ is the bosonic annihilation field operator.

\vspace{3mm}
\textit{Technical remarks:}
In the method outlined above we sample different configurations of the system, using a random walk in the space of configurations. Performing a single step in this walk is well optimized. To collect a good representation of the ensemble one needs to store configurations that are not closely correlated and we only consider samples separated further than the correlation length in the Metropolis random walk.  However, for a large number of atoms, a single step does not change the system significantly and the correlation length becomes very long.
Thus, the computation for $N=10^5$ non-interacting atoms in a spherical trap takes 6 weeks on a multi-core computational server~\footnote{The code is parallelized, and we run it on a computational unit belonging to the Polish infrastructure of supercomputers PL-GRID.}. For higher aspect ratios and atom numbers the computation becomes even more demanding, rendering calculations unfeasible.

We do not account directly for any superposition of Fock states. Our analysis is devised to work for the ideal gas and for a weakly interacting system. We will present a detailed analysis of the method, with benchmarks and generalizations elsewhere.

\begin{figure}[b]
	\centering
	\includegraphics[scale=1]{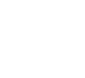}
	\label{fig:Fluctuations3D}
\end{figure}